# Does a dissolution-precipitation mechanism explain concrete creep in moist environments?

*Isabella Pignatelli ([1]), Aditya Kumar ([1]), Rouhollah Alizadeh ([2]), Yann Le Pape ([3]), Mathieu Bauchy ([4*]), Gaurav Sant ([1*], [5])*

**Abstract**

Long-term creep (i.e., deformation under sustained load) is a significant material response that needs to be accounted for in concrete structural design. However, the nature and origin of creep remains poorly understood, and controversial. Here, we propose that concrete creep at RH (relative humidity) ≥ 50%, but fixed moisture-contents (i.e., basic creep), arises from a dissolution-precipitation mechanism, active at nanoscale grain contacts, as is often observed in a geological context, e.g., when rocks are exposed to sustained loads, in moist environments. Based on micro-indentation and vertical scanning interferometry experiments, and molecular dynamics simulations carried out on calcium–silicate–hydrates (C–S–H's), the major binding phase in concrete, of different compositions, we show that creep rates are well correlated to dissolution rates – an observation which supports the dissolution-precipitation mechanism as the origin of concrete creep. C–S–H compositions featuring high resistance to dissolution, and hence creep are identified – analysis of which, using topological constraint theory, indicates that these compositions present limited relaxation modes on account of their optimally connected (i.e., constrained) atomic networks.

[1] Laboratory for the Chemistry of Construction Materials (LC[2]), Department of Civil and Environmental Engineering, University of California, Los Angeles, CA, USA
[2] Giatec Scientific, Ottawa, Canada
[3] Oak Ridge National Laboratory, Oak Ridge, TN, USA
[4] Physics of AmoRphous and Inorganic Solids Laboratory (PARISlab), Department of Civil and Environmental Engineering, University of California, Los Angeles, CA, USA
[5] California Nanosystems Institute (CNSI), University of California, Los Angeles, CA, USA

* Corresponding authors: M. Bauchy (bauchy@ucla.edu) and G. Sant (gsant@ucla.edu)

**Introduction**

Second to water, concrete is by far the most used material in the world [1]. Due to the abundance of necessary raw materials, and the inexpensive nature of ordinary portland cement (OPC), the binder used in concrete, this status is unlikely to change in the near future. However, cementitious materials undergo evolutions in structure and properties in time, resulting in shrinkage and creep, especially upon drying or due to imposed loads, which can ultimately result in fracture and failure [2]. Creep manifests as a time-dependent volume change under sustained load, which can cause dramatic deformations in infrastructure [3]. However, the physical origin of creep in concrete remains poorly understood [4]. It is thought that the creep of concrete, particularly at later ages, is largely caused by the viscoelastic and viscoplastic behavior of the calcium–silicate–hydrate (C–S–H) [5], the major binding phase in concrete, formed when cement reacts with water [6]. While secondary cementitious phases can show viscoelastic behavior [7–9], the rate and extent of deformation of such phases is far less significant than that of the C–S–H [5]. As such, understanding and unambiguously elucidating the mechanism of C–S–H creep is of primary importance in the context of predicting and limiting delayed deformations under load. Knowledge of such mechanism would allow one to understand the relationship between creep and binder composition – a potential route to more effectively utilize concrete.

Different models of C–S–H creep have been proposed, although it is generally acknowledged that none of the existing models are able to explain all the experimental observations [10]. This suggests that the mechanism of creep may change in relation to environmental conditions (i.e., temperature, relative humidity (RH), applied load) or that several creep mechanisms may operate simultaneously. Popular models of basic creep in C–S–H are based on shear deformations of the C–S–H grains [11], consolidation due to seepage and redistribution of pore water under stress [12], rearrangements of the C–S–H gel manifesting in the form of rupture of inter-granular bonds and rearrangements under relative motion [13], rearrangements of C–S–H grains (under load) following a granular medium analogy [14], and/or optimization of the alignments of the C–S–H layers [5]. These models fundamentally rely on idea that the sliding of C–S–H grains or layers, with respect to each other enables a global structural reorganization.

While we do not contest the premise of structural reorganization (e.g., see [15]), we propose an alternative origin and assess the hypothesis of a dissolution-precipitation mechanism for the creep of C–S–H in moist environments (broadly, RH ≥ 50%). This creep mechanism, well known in geology, contributes to upper-crustal deformations [16], and has been observed in aggregates of rock salt immersed in brine at room temperature [17]. The physical mechanism involves the following steps: (a) the external load induces local compressive stresses between adjacent mineral (in this case, C–S–H) grains, (b) this stress provokes a change of the chemical potential of the solid, (c) according to Le Chatelier's principle, the chemical equilibrium between solid and liquid is shifted in order to enhance solubility and reduce stress, (d) the dissolved species diffuse through the water film to lower concentration regions following Fickian diffusion, and (e) this results in the formation of a precipitate, similar (if not equivalent) to the dissolved phase, on the surfaces of unstressed grains. This process leads to a plastic strain in order to accommodate the stress. This series of events persists as long as the local stress exists. Such a process has recently been reported to control creep in gypsum plaster [18] and has been suggested to potentially also apply to hydrated cement solids [18]. According to this model, under appropriate moisture levels, creep would be a consequence of the dissolution of C–S–H in inter-granular water (i.e., including capillary water, and some of the gel water [15]) in high-stress regions. Based on an original approach combining micro-indentation studies of mechanical properties, vertical scanning interferometry based analysis of aqueous dissolution rates of C–S–H, and molecular dynamics simulations of C–S–H's of varying compositions, we identify a significant correlation between creep and dissolution, which strongly supports the hypothesis of such a dissolution-precipitation origin of concrete creep.

**Results**
**Creep propensity of C–S–H with respect to Ca/Si**
In order to demonstrate or refute the hypothesis of a dissolution-precipitation mechanism controlling the creep of C–S–H, we studied by experiments and simulations the composition dependence of creep and dissolution in C–S–H. Monophasic C–S–H of different compositions (with a Ca/Si molar ratio of 1.2 and 1.5) were synthesized from the pozzolanic reaction between CaO and amorphous $SiO_2$ (see the Supplemental Material [19]). C–S–H solids with different packing fractions $\varphi$ were prepared by compaction under a pressure $P$ ($\varphi$ being a function of $P$), and their resistance to creep was tested by micro-indentation. As reported elsewhere [9], we assume that the C–S–H compacts are representative of the C–S–H phase in hardened cement paste, i.e., showing similar elastic modulus, hardness and creep modulus correlations to their packing fraction. Note that, herein, we focus on basic creep, that is, creep occurring without a simultaneous change in the internal moisture state. As such during the indentation analysis, the relative humidity of pre-conditioned C–S–H samples was kept constant at 11% RH during testing.

We acknowledge that this low RH corresponds to the existence of a single monolayer of water (i.e., a statistical monolayer) on the C–S–H surfaces, and presence of the full amount of interlayer water inside the solid C–S–H [15]. For these RH conditions, significant dissolution of the C–S–H is unlikely to occur. However, atomistic simulation results have revealed, for RHs as low as 7%, the interstitial space between two surfaces of C–S–H separated by 10 Å was almost fully occupied by adsorbed water molecules [20]. This suggests that, although there is a single monolayer of water on the C–S–H surface on average, the amount of water can be far more significant at points of contact between the C–S–H grains, that is, where stresses would be the largest – and where dissolution is potentially most favored. While arguably, dissolution (and hence creep) would be significantly enhanced, and favored when liquid water is present – creep is dictated by a composition-driven mechanical instability of the C–S–H phases as captured by rigidity theory (see below). As noted in the discussion, this atomistic instability also favors, or hinders the dissolution of a given C–S–H composition, vis-à-vis, others.

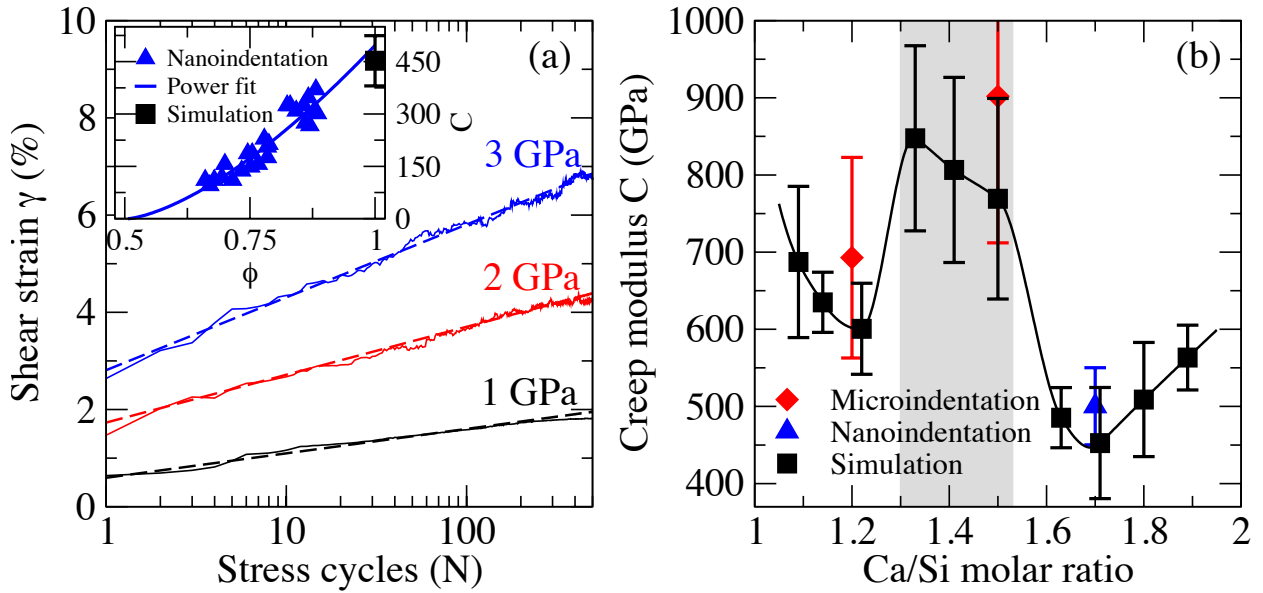

**(Color online) Figure 1:** Logarithmic creep of solid calcium–silicate–hydrate (C–S–H). (a) Shear strain γ of solid C–S-H (Ca/Si = 1.7) with respect to the number of applied stress perturbation cycles, when subjected to 1, 2, and 3 GPa shear stresses τ. The dashed lines indicate logarithmic fits following $\gamma = (\tau/C) \log(1 + N/N_0)$, permitting to evaluate the creep modulus *C*. The inset shows *C* with respect to packing fraction φ, as obtained by nano-indentation [14]. The values are fitted by a power law $C = A (\varphi - 0.5)^\alpha$ and extrapolated to zero porosity to be compared with the value obtained by the present simulations. (b) Computed creep modulus of solid C–S–H with respect to the Ca/Si molar ratio. The values are compared with experimental measurements obtained by micro-indentation [9] and nano-indentation [14]. The grey area indicates the extent of the compositional window in which a maximum resistance to creep is observed.

Micro-indentation creep tests carried out on C–S–H compacts show that under a constant indentation load, we observe a logarithmic increase of the penetration depth (see Reference [9] for more details), which, in agreement with previous nano-indentation experiments [14], suggests that the creep of C–S–H is intrinsically logarithmic with the time. Following a well-established methodology [9] (see the Supplemental Material [19]), indentation hardness *H*, indentation modulus *M*, and creep modulus *C* were quantified. Note that these quantities depend on the packing fraction φ [14]. In order to evaluate the dependence of *C* on the composition (i.e., Ca/Si molar ratio) of C–S–H without any contribution from the porosity, the results were extrapolated to zero porosity. However, due to the lack of accuracy in assessing φ precisely, direct extrapolation is challenging. On the other hand, for a given C–S–H composition, *C* features a strongly linear correlation with *H*. The zero-porosity hardness $H_0$ of C–S–H samples of varying Ca/Si molar ratios was recently calculated by atomistic simulations and showed excellent agreement with experimental data [21]. As such, by fitting *C* versus *H* and extrapolating *C* to $H = H_0$, one can estimate the creep modulus at zero porosity, which is shown in Figure 1b, along with data obtained via nano-indentation for Ca/Si = 1.7 [14]. We note that C–S–H with Ca/Si = 1.5 exhibits the highest resistance to creep, with a creep modulus around 80% higher than that obtained for Ca/Si = 1.7. Such non-linear behavior is very different from that of indentation hardness and creep modulus, both of which decrease monotonically with Ca/Si [21]. The details of the indentation study and the *C(H)* linear curves that permit extrapolation of *C* to zero porosity can be found in Reference [9].

**Molecular dynamics simulations of C–S–H creep**

To obtain more detailed insight into the physical mechanism of creep in C–S–H, molecular dynamics (MD) simulations were carried out. The atomistic models of C–S–H from Pellenq *et al.* [21, 22] were used as an initial starting point as their computed mechanical properties (i.e., indentation modulus and hardness) have been found to be in excellent agreement with experimental data [21]. Although fine structural details of this model have been criticized [23–25], broadly, and to the best of our knowledge, it remains the only model that is capable of describing C–S–H compositions across a wide range of Ca/Si molar ratios. In addition, the main outcomes of this study do not depend on the fine structural details of the chosen model. Although MD simulations are typically limited to a few nanoseconds, we developed for this study, a novel technique allowing us to simulate creep over longer timescales. This technique involves applying a constant shear stress $\tau$, mimicking experimental measurements of deviatoric creep [26]. In addition to $\tau$, periodic, small stress perturbations [27] are then applied (in cycles, see the Supplemental Material [19]). Each perturbation cycle slightly deforms the energy landscape experienced by the system, thereby permitting jumps over select energy barriers, allowing the system to relax towards lower energy states.

Careful analysis of the internal energy shows that the height of the energy barriers, through which the system transits across each cycle, remains roughly constant over successive cycles. On the basis of transition state theory, which states that the time needed for a system to jump over an energy barrier $E_A$ is proportional to $\exp(-E_A/kT)$, we can assume that each cycle corresponds to a constant duration $\Delta t$, so that a fictitious time can be defined as $t = N\Delta t$, where $N$ is the number of cycles [28]. Note that these transitions occur spontaneously, although the duration $\Delta t$ before each jump is too long to be accessible from conventional atomistic simulations. We recently used a similar accelerated relaxation method to study room-temperature relaxation of silicate glasses [29], and showed that the dynamics of relaxation does not depend on the amplitude of the stress perturbations, as long as they remain significantly lower than the yield point of the material. Figure 1a shows the computed shear strain $\gamma$ of a C–S–H system with Ca/Si = 1.7, with respect to the number of stress perturbation cycles $N$, for different constant stresses $\tau$ = 1, 2, and 3 GPa. In agreement with experimental indentation profiles, we observe a logarithmic creep response, which can be expressed as:

$$\gamma(N) = (\tau/C)\log(1 + N/N_0) \qquad (1)$$

where, $N_0$ is a fitting parameter and *C* is the creep modulus which is determined by fitting the computed shear strain with respect to the number of stress cycles (see Figure 1a). Interestingly, we find that the computed shear strains are proportional to the applied constant shear stress $\tau$. As such, *C* does not depend on the applied stress and, thereby, appears to be an intrinsic property of the material. We observe, however, that this holds only so long as the applied stress remains lower than the yield stress of the sample [27]. Note that, as our simulations do not consider any porosity, the computed values of *C* can only be compared to experimental values extrapolated to zero porosity. As shown in the inset of Figure 1a, the obtained *C* (≈450 GPa) is in very good agreement with nano-indentation data [14], extrapolated to a packing fraction of 1. To the best of our knowledge, this is the first time that the creep propensity (as indicated by the creep modulus) of cementitious, or other viscoelastic materials has been successfully reproduced by atomistic simulation. This approach was applied to other C–S–H compositions to better understand the relationship between their composition and resistance to creep. As shown in Figure 1b, the computed *C* values show a non-linear evolution with the Ca/Si, which manifests in the form of a broad maximum around Ca/Si = 1.5. Once again, the obtained *C* values are in excellent agreement with micro-indentation data extrapolated to zero porosity.

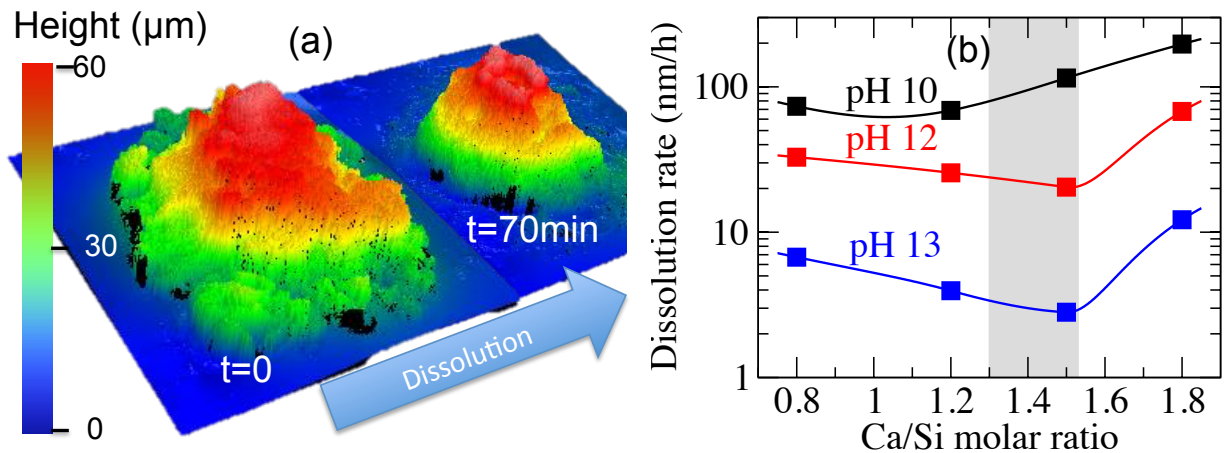

**(Color online) Figure 2:** Dissolution of calcium–silicate–hydrate (C–S–H). (a) Illustration of C–S–H (Ca/Si = 0.8, pH = 10) dissolution as visualized using vertical scanning interferometry (VSI) at zero time and after 70 min of solvent contact time. Dissolution is tracked by measuring the decrease in the grains height over time. (b) Dissolution rates (25°C, 1 bar) of C–S–H samples with the respect to the Ca/Si molar ratio, at different pH. The grey area indicates the extent of the compositional window in which a maximum resistance to creep and dissolution is observed (see Figures 1 and 3).

**Dissolution sensitivity of C–S–H with respect to Ca/Si**
We now investigate whether a dissolution-precipitation sensitivity of creep, as a function of composition, would explain the maximum resistance to creep observed around Ca/Si = 1.5. In the dissolution-precipitation model, critical variables include the structure of the inter-grain contacts, grain packing and their size distributions, and identifying the slowest (i.e., rate controlling) step of the process. Due to the small spacing between inter-granular contacts (on the order of nanometers [20]), the transport of solubilized ions between grains will be fast under conditions of sufficient moisture and where transport confinement is slight, which suggests that the dissolution rate (i.e., rather than ionic diffusion) is rate-controlling. According to Raj's model [17], the strain rate should then be proportional to: (1) the applied stress and (2) the dissolution rate (constant), and (3) inversely proportional to the grain size – though given the similar nanoscale dimension of C–S–H grains [30], size is potentially less relevant. Since (1) has already been validated in the previous section, we evaluated the dissolution kinetics of C–S–H to evaluate its validity, and relevance to creep behaviors.

The dissolution rates of C–S–H powders, equivalent to those that formed the compacts used in micro-indentation testing (with Ca/Si = 0.8, 1.2, 1.5, and 1.8) were measured using vertical scanning interferometry [31] (VSI, see the Supplemental Material [19]). The VSI technique has been extensively applied to measure the dissolution rates of minerals of geological and technological relevance [32, 33]. By directly tracking the evolution of the surface topography in time, with sub-nanometer vertical resolution, VSI accesses the "true" dissolution rate of a dissolving solid. Unlike dissolution assessments that are based on analysis of solution compositions, which may be affected by aspects including metastable barrier formation, incongruency in dissolution or, ion adsorption, VSI analytics are not influenced by such complexities. Dissolution rates of the C–S–H solids were quantified using a rain-drop procedure [31], wherein both the solution pH and composition (i.e., the under-saturation level with respect to the dissolving solid) are kept constant over the course of the experiment (see Figure 2a). Dissolution experiments were conducted at room temperature and at pH 10, 12, and 13, the latter value corresponding to the typical pH of the pore-fluid in mature cement pastes [30].

Figure 2b shows the dissolution rates of the various C–S–H compositions as a function of the Ca/Si molar ratio, for different solvent pH's. Overall, we note that, for every C–S–H composition, the dissolution rate decreases with pH, as also observed for limestone, alite, ordinary portland cement, and gypsum – since the under-saturation level, described to the first order by the hydroxyl ion activity, increases (i.e., becoming a larger number, with a negative sign) accordingly [31]. In addition, at constant pH, we observe a strongly non-linear evolution of the dissolution rate with respect to the Ca/Si. Interestingly, at pH 12 and 13, a minimum of dissolution is observed for Ca/Si = 1.5, that is, for the range of compositions featuring an increased resistance to creep (see the previous section). This minimum shifts to a lower Ca/Si ratio at pH 10. This is a significant observation, which as shown in the inset of Figure 3a, indicates a strong correlation between dissolution and creep rates, which supports the hypothesis of a dissolution-precipitation mechanism as the origin of concrete creep in moist environments – i.e., those containing water with bulk or *bulk-like* properties, in pores exposed to the atmosphere (ambient RH), or ink-bottle/dead-end pores – wherein bulk water may remain shielded by smaller pore entryways – and would hence not deplete until lower RHs [15].

**Discussion**
Based on the atomistic simulations detailed above, we aim to identify the common structural origin of the maximum resistance to creep and dissolution observed around Ca/Si = 1.5. To this end, we used topological constraint theory [34–36] (TCT), a framework which has been successfully applied to understand compositional controls on the properties of glasses [37–39]. TCT captures the topology of atomic networks while filtering out less relevant structural details that ultimately do not affect the macroscopic response or properties. Within the TCT framework, atomic networks are described as mechanical trusses, in which the atoms experience displacement constraints, as imposed by the radial and angular chemical bonds. Therefore, following Maxwell's stability criterion [40], an atomic network is described as flexible, stressed-rigid, or isostatic, if the number of topological constraints per atom ($n_c$) is lower, higher, or equal to three, i.e., the number of degrees of freedom per atom in three dimensions (see Figure 3a).

Our use of TCT is stimulated by the observation that isostatic glasses tend to show weak aging over time [41, 42], a feature potentially relevant to C–S–H, and hence concrete creep. Recently, we extended TCT to handle poorly crystalline materials like C–S–H [43]. Interestingly, as shown in Figure 3b, we note that C–S–H features a rigidity transition at Ca/Si = 1.5 [37], while being stressed-rigid ($n_c > 3$) at low Ca/Si and flexible ($n_c < 3$) at higher Ca/Si. As such, as shown in Figure 3c, we observe that a maximum resistance to creep and dissolution is achieved when the atomic network is isostatic ($n_c = 3$) at Ca/Si = 1.5. This constitutes, to the best of our knowledge, the first quantitative evidence of a link between atomic topology (and hence composition), and resistance to creep and dissolution. To explain these results, we propose the following atomistic view. (1) Due to a lack of *mechanical-equivalent* constraints, flexible atomic networks feature internal floppy modes of deformation [35]. These low-energy modes facilitate the reorganization of the network under stress (for example, see recent experimental evidence for the reorganization of C–S–H in cement paste, under the imposition of a hydrostatic drying stress [15]), thereby enhancing creep. Similarly, lower network connectivity ($n_c < 3$) has been shown to enhance dissolution [44]. (2) On the other hand, stressed-rigid atomic networks show internal eigenstresses [45], which originates from a steric frustration of the network. Indeed, due to their high number of chemical bonds, some constraints become redundant, that is, all of them cannot be satisfied simultaneously – just as it would be impossible to adjust one of the angles of a rigid triangle with three edges fixed. These internal stresses induce instabilities in the network and, thereby, act as a driving force for phase separation or devitrification of glasses [36, 46]. Such instability, represented by an increase in entropy, also enhances dissolution sensitivity. Note that such instabilities will only be observed in disordered materials, as crystals, on account of their optimized geometry, have a structure that does not induce any internal stress – in absence of an external stimulus. In others words, even if the number of topological constraints inside a crystal becomes larger than the number of degrees of freedom, all the constraints are perfectly compatible with each other, so that no internal frustration is created. Coming back to creep, stressed-rigid glasses have been shown to feature low elastic recovery after pressurization (densification) [47]. Indeed, once placed under pressure, due to their high number of constraints, the network remains permanently locked in its densified state[48]. If creep is seen as a succession of small cycles as stress, in stressed-rigid C–S–H compositions, each cycle will induce an irreversible deformation (compaction in the case of compression, and elongation in the case of tensile loading), thereby resulting in significant accumulative deformations with longer imposed stress durations, and, as a result, a lower resistance to creep.

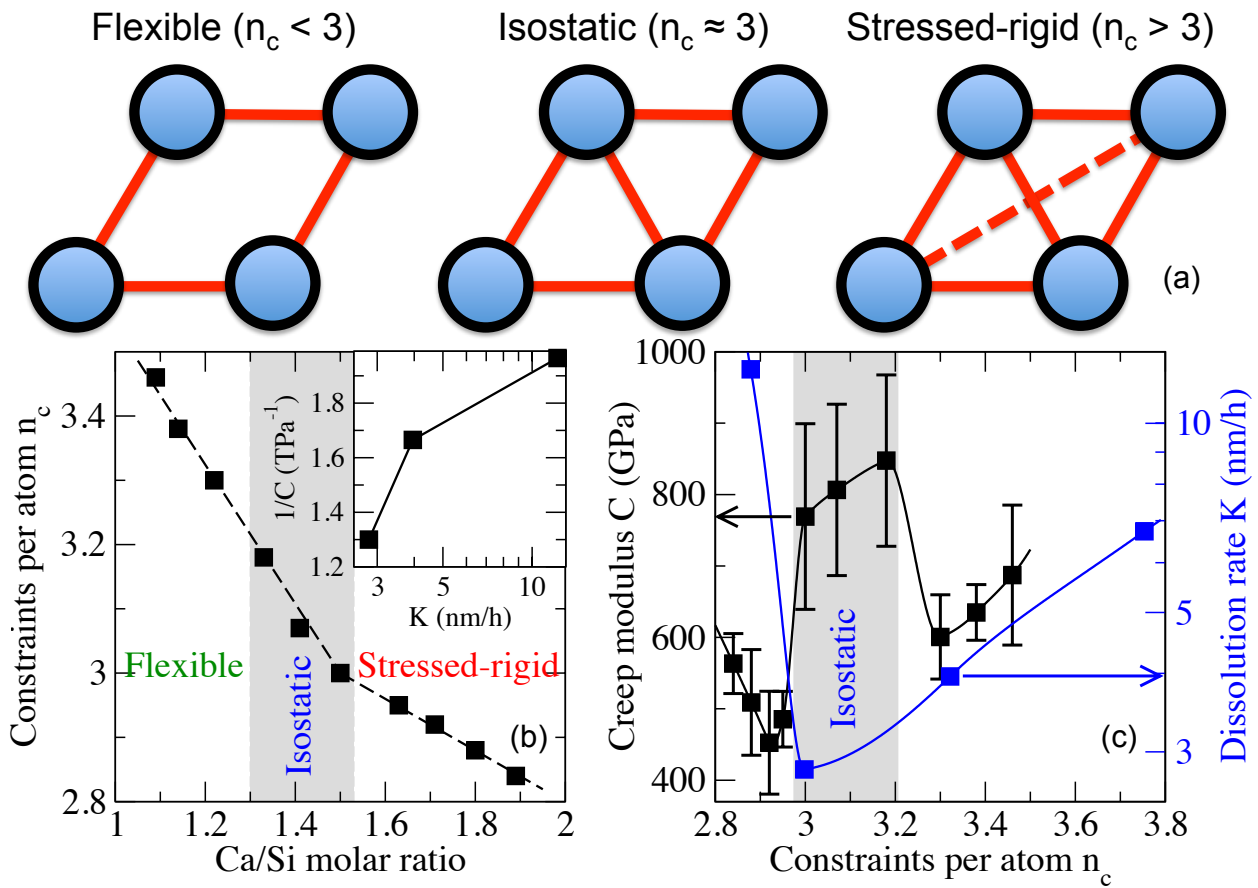

**(Color online) Figure 3:** Topological analysis of calcium–silicate–hydrate (C–S–H). (a) The three states of rigidity of a mechanical truss. Flexible networks feature internal degrees of freedom, stressed-rigid ones show eigen-stress, whereas isostatic ones are rigid but free of internal stress. The dashed line indicates a frustrated constraint, which cannot be satisfied due to redundant constraints, thereby resulting in eigen-stress. (b) Number of chemical topological constraints per atom $n_c$ in C–S–H with respect to the Ca/Si molar ratio. The dashed line is a guide for the eye. The grey area indicates the extent of the compositional window in which a maximum resistance to creep and dissolution is observed (see Figures 1 and 2), also corresponding to the range of isostatic compositions ($n_c \approx 3$), effectively separating the flexible ($n_c < 3$) from the stressed-rigid domains ($n_c > 3$). The inset shows the correlation between the inverse of the creep modulus $C$ and the dissolution rate $K$ of C–S–H. The line is a guide for the eye. (c) Creep modulus $C$ and dissolution rate $K$ (pH 13) of C–S–H with respect to the number of topological constraints per atom $n_c$. The grey area indicates the position of the isostatic window (see text).

Alternatively, these results can be understood by considering the roughness of the energy landscape, which is determined by both the bond and the floppy modes densities [49]. Indeed, the bond density, that is, the connectivity, tends to induce the creation of energy basins. In parallel, the floppy modes density leads to the formation of channels between these basins, therefore extending the possibilities of relaxation for the flexible networks, and, thereby, enhancing creep and dissolution. However, for the stressed-rigid networks, the frustration induced by unsatisfied constraints acts as an elastic energy [50], which, in turn, induces some jumps between the basins, which, again, extends the possibilities of relaxation. As such, isostatic networks do not feature any barrier-less channels between the basins and do not possess any eigenstress-induced driving force to jump over the energy barriers. Such optimally constrained networks therefore show limited modes of relaxation, which render them resistant to creep and stable with respect to dissolution.

**Conclusions**
By simultaneously quantifying the propensity of C–S–H compositions to dissolve and to creep, we have clarified that dissolution and creep are strongly correlated to, and indicative of each other. Such correlation offers the first direct evidence that a dissolution-precipitation mechanism could be the primary origin of the concrete creep in moist environments, as has been observed for geological materials. More generally, we have shown that atomic topology is a fundamental variable that renders a material, sensitive or not, to aging and dissolution. This is significant as it provokes an opportunity for tuning material compositions. On this basis, we find that shifting the composition of the binding C–S–H phase to lower Ca/Si would increase the resistance to creep of concretes. It should be noted however, that these conclusions only refer to C–S–H compositions whose moisture content is fixed (i.e., no drying occurs). However, even in this case, the imposition of a stress would result in: (i) an internal redistribution of moisture and (ii) a change in the internal stress that facilitates dissolution as the body creeps [51]. Depending on how this occurs – as moisture distributes internally, dissolution/precipitation (and hence creep) could increase in moisture rich zones, which can support more dissolution, and decrease in drier regions. On the other hand, since creep induces stress relaxations, the influence of stress on C–S–H dissolution would lift locally, following creep – resulting in reduced dissolution. Effects of this nature could result in spatial variations in creep – until the dominant process ensures that globally, creep kinetics are altered. Actions of this nature, when combined with the effects of drying (i.e., mass change, resulting in drying creep) and mechanical loading may explain the rate of creep deformations seen in practice, and explain behaviors such as the Pickett effect that is seen in cementing systems [52–54], and have yet not been fully explained.

**Acknowledgements**
The authors acknowledge full financial support for this research provided by: The U.S. Department of Transportation (U.S. DOT) through the Federal Highway Administration (DTFH61-13-H-00011), the National Science Foundation (CAREER Award: 1235269), The Oak Ridge National Laboratory operated for the U.S. Department of Energy by UT-Battelle (LDRD Award # 4000132990), and the University of California, Los Angeles (UCLA). Access to computational resources was provisioned by the Physics of AmoRphous and Inorganic Solids Laboratory (PARISlab), the Laboratory for the Chemistry of Construction Materials (LC$^2$) and the Institute for Digital Research and Education (IDRE) at UCLA. This research was conducted in: Laboratory for the Chemistry of Construction Materials (LC$^2$) and Physics of AmoRphous and Inorganic Solids Laboratory (PARISlab) at UCLA. The authors gratefully acknowledge the support that has made these laboratories and their operations possible. The contents of this paper reflect the views and opinions of the authors, who are responsible for the accuracy of the datasets presented herein, and do not reflect the views and/or policies of the funding agencies, nor do the contents constitute a specification, standard or regulation. This manuscript has been co-authored by the Oak Ridge National Laboratory, managed by UT-Battelle LLC under Contract No. DE-AC05-00OR22725 with the U.S. Department of Energy. The publisher, by accepting the article for publication, acknowledges that the U.S. Government retains a nonexclusive, paid-up, irrevocable, worldwide license to publish or reproduce the published form of this manuscript, or allow others to do so, for U.S. Government purposes. The Department of Energy will provide public access to these results of federally sponsored research in accordance with the DOE Public Access Plan (http://energy.gov/downloads/doe-public-access-plan).